\documentclass[12pt, a4paper] {article}

\usepackage{amsfonts}
\usepackage{verbatim}
\usepackage{color}

\begin{document}

\begin{center}
{\bf \large BRST analysis of \\topologically massive gauge theory: novel observations}\\

\vskip 2.5cm

{\bf R. Kumar$^{(a)}$, R. P. Malik$^{(a, b)}$}\\
{\it $^{(a)}$Physics Department, Centre of Advanced Studies,}\\
{\it Banaras Hindu University, Varanasi - 221 005, (U.P.), India}\\

\vskip 0.1cm

{\bf and}\\

\vskip 0.1cm

{\it $^{(b)}$DST Centre for Interdisciplinary Mathematical Sciences,}\\
{\it Faculty of Science, Banaras Hindu University, Varanasi - 221 005, India}\\
{\small {e-mails: raviphynuc@gmail.com, malik@bhu.ac.in}}
\end{center}

\vskip 2cm

\noindent
{\bf Abstract:} A dynamical non-Abelian 2-form gauge theory (with $B \wedge F$ term) is endowed
with the ``scalar'' and  ``vector'' gauge symmetry transformations. In our present endeavor,
we exploit the latter gauge symmetry transformations and perform the Becchi-Rouet-Stora-Tyutin (BRST) 
analysis of the four (3 + 1)-dimensional (4D) topologically massive non-Abelian 2-form gauge theory. We 
demonstrate the existence of some novel features that have, hitherto, not been observed in the 
context of BRST approach to 4D (non-)Abelian 1-form as well as Abelian 2-form and 3-form gauge theories. 
We comment on the differences between the novel features that emerge in the BRST analysis of 
the ``scalar'' and ``vector'' gauge symmetries.

\vskip 1.5 cm

\noindent
PACS \hskip .3cm {\sf 11.15.Wx} $-$ Topologically massive gauge theories\\
PACS \hskip .3cm {\sf 11.15.-q} \hskip .22cm $-$ Gauge field theories\\

\noindent
{\it Keywords:} Dynamical non-Abelian 2-form theory; topological $(B \wedge F)$ term;
``vector'' gauge symmetries; (anti-)BRST symmetries; Curci-Ferrari type restrictions;
nilpotency and anticommutativity

\newpage
\noindent
{\bf \large 1. Introduction}\\

\noindent
In recent years, there has been a renewed interest in the study of 4D topologically massive (non-)Abelian 
gauge theories. In such theories, there is an explicit coupling between the 1-form and 2-form gauge fields through 
the celebrated 
$(B \wedge F)$ term. In fact, it has been 
shown that the 1-form gauge field acquires a mass, in a very natural fashion [1], for
the above 4D topologically massive gauge theories. As a result, the above models [1-5] 
provide an alternative to the Higgs mechanism of the standard model of high energy physics as
far as the mass generation of the 1-form gauge field is concerned.

In the above context, it may be mentioned that we have carried out the BRST analysis 
[6,7] of the 4D 
{\it Abelian} 2-form theory (with topological mass term) and obtained the off-shell nilpotent and absolutely
 anticommuting (anti-)BRST
transformations by exploiting the geometrical superfield approach [8,9]. The latter formalism 
has also been applied to the 4D dynamical {\it non-Abelian} 2-form theory 
(with celebrated topological $B \wedge F)$ term)
and we have exploited its ``scalar" and ``vector" gauge transformations (see, $e.g.,$ [10]) to derive the 
appropriate Lagrangian densities as well as proper (nilpotent and anticommuting)
(anti-)BRST symmetry transformations.

In a very recent paper [11], we have performed the BRST analysis of the 4D {\it non-Abelian} topologically 
massive theory and shown that the 
conserved and nilpotent (anti-)BRST charges, corresponding to the ``scalar" gauge symmetry of the theory, 
are unable to generate the (anti-)BRST transformations 
corresponding to the $B_{0i}$ component of the antisymmetric tensor gauge field
$B_{\mu\nu}$ (with  $B_{\mu\nu} = B_{\mu\nu} \cdot T$) and the 1-form ($K^{(1)} = dx^\mu K_\mu \cdot T$) auxiliary field $K_\mu$
(with  $K_\mu = K_\mu \cdot T$). This happens to be a novel observation in [11].

The central theme of our present paper is to exploit the ``vector" gauge symmetry 
transformations of the above topologically massive non-Abelian theory and explore its details 
within the framework of BRST 
formalism. As it turns out, we observe yet another novel feature in the BRST analysis of the above 
topologically massive gauge theory. 
We find that the conserved and nilpotent (anti-)BRST charges are not able to generate the 
proper (anti-)BRST symmetry
transformations {\it only} for the auxiliary field $K_\mu$. 
This is a new result that is quite different from the BRST analysis of the 4D (non-)Abelian 1-form [12,13] and 
Abelian 2-form and 3-form gauge theories [6,7,14]. We lay emphasis on the fact that the novel features, 
from the BRST analysis of ``scalar"
and ``vector" gauge symmetry transfromations
 of the {\it same} non-Abelian topologically massive theory, are quite different.

The material of our present paper is organized as follows. In our second section, we recapitulate 
the bare essential of the ``vector" gauge symmetry transformations and derive the generator 
corresponding to them. In section three, we discuss the BRST symmetries corresponding to the above 
``vector" gauge symmetry transformations and deduce the BRST charge. Our section four is devoted 
to the derivation of anti-BRST symmetries and corresponding nilpotent and conserved charge. 
We deal with the ghost symmetries in section five where we also deduce the BRST
algebra. Finally, we discuss our main results and make a few concluding remarks in section six.

In our Appendix, we briefly comment on the St{\"u}ckelberg formalism, Abelian
Higgs model and our present model of the dynamical non-Abelian 2-form gauge theory
(i.e. a model of topologically massive gauge theory).  \\

\noindent
{\bf \large 2. Preliminaries: symmetry transformations}\\

\noindent
Let us begin with the Lagrangian density of the 4D dynamical non-Abelian 2-form 
gauge theory\footnote{We adopt the conventions and notations such that the background 
Minkowaskian 4D spacetime manifold is endowed with the flat metric $\eta_{\mu\nu} =$ 
diag $(+1, -1, -1, -1).$ This entails upon two non-null vectors $A_\mu$ and $B_\mu$ to have: 
$A_\mu B^\mu = \eta_{\mu\nu} A^\mu B^\nu = A_0 B_0 - A_i B_i$ where the
Greek indices $\mu, \nu, \eta, ... = 0, 1, 2, 3$ and Latin indices $i, j, k, ...= 1, 2, 3.$ 
We choose here the 4D Levi-Civta tensor $\varepsilon^{\mu\nu\eta\kappa}$ to obey 
$\varepsilon^{\mu\nu\eta\kappa}\varepsilon_{\mu\nu\eta\kappa} = - 4!,\;
\varepsilon^{\mu\nu\eta\kappa}\varepsilon_{\mu\nu\eta\sigma} = - 3! \;\delta^\kappa_\sigma, etc.,$ 
and $\varepsilon_{0123} = + 1 = - \varepsilon^{0123}.$ In the $SU(N)$ algebraic space, we follow 
$P \cdot Q = P^a \;Q^a$ and $(P\times Q)^a = f^{abc}\; P^b\; Q^c$ where $a, b, c, ... = 1, 2, 3, ... 
....(N^2 - 1)$.}, that incorporates the celeberated $(B\wedge F)$ term with the 
mass parameter $m$, as given below (see, $e.g.,$ [3-5,11])
\begin{eqnarray}
{\cal L}_{(0)} = - \; \frac{1}{4} \; F^{\mu\nu} \cdot F_{\mu\nu} 
+ \frac {1}{12} \; H^{\mu\nu\eta} \cdot  H_{\mu\nu\eta} 
+ \frac {m}{4} \; \varepsilon^{\mu\nu\eta\kappa} \; B_{\mu\nu} \cdot F_{\eta\kappa}.  
\end{eqnarray}
In the above, the curvature tensor $F_{\mu\nu} = \partial_\mu A_\nu - \partial_\nu A_\mu 
- (A_\mu \times A_\nu),$ corresponding to the 1-form
$(A^{(1)} = dx^\mu A_\mu \cdot T)$ non-Abelian field $A_\mu,$ has been derived from the curvature 2-form 
$F^{(2)} = dA^{(1)} + i (A^{(1)} \wedge A^{(1)}) \equiv \frac {1}{2!} \; (dx^\mu \wedge dx^\nu) F_{\mu\nu}.$ 
In exactly similar fashion, the curvature 3-form $H^{(3)} = \frac {1}{3!} \; 
(dx^\mu \wedge dx^\nu \wedge dx^\eta) H_{\mu\nu\eta}$  defines the  totally 
antisymmetric third-rank tensor\footnote{It is possible to eliminate the auxiliary field $K_\mu$ from 
$H_{\mu\nu\eta}$ by the shift transformation $B_{\mu\nu} \to \tilde B_{\mu\nu} + (D_\mu K_\nu - D_\nu K_\mu)$.
We comment on, some aspects of it, in our Appendix.}
\begin{eqnarray}
H_{\mu\nu\eta} &=& (\partial_\mu B_{\nu\eta} + \partial_\nu B_{\eta\mu} + \partial_\eta 
B_{\mu\nu})- \bigl [(A_\mu \times B_{\nu\eta}) + (A_\nu \times B_{\eta\mu})\nonumber\\
&+& (A_\eta \times B_{\mu\nu}) \bigr ] - \bigl [(K_\mu \times F_{\nu\eta}) 
+ (K_\nu \times F_{\eta\mu}) + (K_\eta \times F_{\mu\nu}) \bigr ],
\end{eqnarray}
in terms of the 1-form $(K^{(1)} = dx^\mu K_\mu \cdot T)$ auxiliary field $K_\mu$,  
2-form $(B^{(2)} = \frac {1}{2!} (dx^\mu \wedge dx^\nu) B_{\mu\nu} \cdot T)$ antisymmetric tensor 
gauge field $B_{\mu\nu}$ and the antisymmetric curvature tensor $F_{\mu\nu}$ 
(defined earlier). Here the $SU(N)$ 
generators $T^a$  $(a = 1, 2.... N^2 - 1)$ satisfy the Lie algebra $[T^a, T^b] = i f^{abc}\; T^c$ 
where the structure constants $f^{abc}$ can be chosen to be totally antisymmetric in 
$a, b$ and $c$ for the semi-simple Lie group $SU(N)$ under consideration (see, $e.g.,$ [13]).

The above Lagrangian density (1), for the 4D topologically massive non-Abelian gauge theory, 
respects the following infinitesimal and continuous ``vector" gauge symmetry\footnote{It is 
straightforward to check that ${\cal L}_{(0)}$ also respects ($i.e. \; \delta_{gt} {\cal L}_{(0)} = 0 $) 
the ``scalar" gauge transformations $(\delta_{gt})$ corresponding to the usual 1-form non-Abelian 
gauge field: $\delta_{gt} A_\mu = D_\mu \Omega,\; \delta_{gt} B_{\mu\nu} = - (B_{\mu\nu} \times \Omega), 
\delta_{gt} K_\mu = - (K_\mu \times \Omega),\delta_{gt} F_{\mu\nu} = - (F_{\mu\nu} \times \Omega),
\delta_{gt} H_{\mu\nu\eta} = - (H_{\mu\nu\eta} \times \Omega)$ where $\Omega 
= \Omega \cdot T \equiv \Omega^a\; T^a$ is the $SU(N)$ valued infinitesimal gauge (Lorentz ``scalar") 
parameter and the 
covariant derivative $D_\mu \Omega = \partial_\mu \Omega - (A_\mu \times \Omega)$ [3,4,11].} 
transformations $(\delta_v)$ [3,4] 
\begin{eqnarray}
&\delta_v A_\mu = 0, \qquad \delta_v B_{\mu\nu} = - (D_\mu \Lambda_\nu - D_\nu \Lambda_\mu), & \nonumber\\
&\delta_v K_\mu = - \Lambda_\mu, \qquad \delta_v F_{\mu\nu} = 0 , \qquad  \delta_v H_{\mu\nu\eta} = 0,&
\end{eqnarray}
because the Lagrangian density (1) transforms, under (3), as  
\begin{eqnarray}
\delta_v {\cal L}_{(0)}  = - \;\partial_\mu \Big[{\frac {m}{2}} 
\; \varepsilon^{\mu\nu\eta\kappa} \; F_{\nu\eta} \cdot \Lambda_\kappa\Big],
\end{eqnarray}
where $\Lambda_\mu = \Lambda_\mu \cdot T$ is an infinitesimal Lorentz ``vector'' gauge parameter for 
the transformations $\delta_v$. It is evident from (3) that the action $S = \int d^4x {\cal L}_{(0)}$  
remains invariant under the ``vector" symmetry transformations (3).

Noether theorem states that the continuous ``vector" gauge transformations (3) would lead to the 
derivation of a conserved current. The precise expression for this current is as follows:
\begin{eqnarray}
J^\mu_{(v)} &=& - H^{\mu\nu\eta} \cdot (D_\nu \Lambda_\eta) 
+ \frac {m}{2}\; \varepsilon^{\mu\nu\eta\kappa} \; F_{\nu\eta} \cdot \Lambda_\kappa.
\end{eqnarray}
It can be checked that $\partial_\mu J^\mu_{(v)} = 0$ if we exploit the following Euler-Lagrange 
(E-L) equations of motion for all relevant fields, namely;  
\begin{eqnarray}
&&D_\mu H^{\mu\nu\eta} = \frac {m}{2} \; \varepsilon^{\nu\eta\kappa\sigma} F_{\kappa\sigma}, 
\qquad (H^{\mu\nu\eta} \times F_{\nu\eta}) = 0,\nonumber\\
&&D_\mu \Big[F^{\mu\nu} + (H^{\mu\nu\eta} \times K_\eta) 
- \frac{m}{2} \; \varepsilon^{\mu\nu\eta\kappa} B_{\eta\kappa}\Big] 
+ \frac {1}{2} \; (H^{\nu\eta\kappa} \times B_{\eta\kappa}) = 0,
\end{eqnarray}
that emerge from the Lagrangian density ${\cal L}_{(0)}.$ The above conserved Noether current 
leads to the derivation of conserved charge $Q_{(v)} = \int d^3x J^0_{(v)}$ as 
\begin{eqnarray}
Q_{(v)} &=& \int d^3x \Big [-  H^{0ij} \cdot \Big( D_i \Lambda_j\Big ) 
+ \frac {m}{2}\; \varepsilon^{0ijk} \; F_{ij} \cdot \Lambda_k \Big]\nonumber\\
&\equiv & \int d^3x \Big [- \frac {1}{2} \;  H^{0ij} \cdot \Big ( D_i \Lambda_j - D_j \Lambda_i \Big ) 
+ \frac {m}{2}\; \varepsilon^{0ijk} \; F_{ij} \cdot \Lambda_k \Big].
\end{eqnarray}
The above conserved charge is the generator of the infinitesimal ``vector" gauge transformations (3). 
It is interesting, however, to point out that it generates {\it only} the following infinitesimal transformations
\begin{eqnarray}
\delta_v B_{ij} = - i [B_{ij},\; Q_{(v)}] = - (D_i \Lambda_j - D_j \Lambda_i), \quad
\delta_v A_\mu = - i [A_\mu,\;Q_{(v)}] = 0,
\end{eqnarray}
which are a part of the total transformations (3) that corresponds to our ``vector" 
gauge symmetry transformations. It can be seen that the transformations $\delta_v B_{0i}$ 
and $\delta_v K_\mu$ are {\it not} generated by the conserved charge $Q_{(v)}.$
The former transformation can be derived by using the precise techniques of BRST formalism. 
We do the same in our next section.\\

\noindent
{\bf \large 3. BRST symmetries and BRST charge}\\

\noindent
We begin with the BRST invariant Lagrangian density ${\cal L}_B$ (which is the generalization 
of the starting Lagrangian density ${\cal L}_{(0)}$ (cf. (1))) such that the gauge-fixing 
and Faddeev-Popov ghost terms are incorporated in it. Such an appropriate (BRST-invariant) 
Lagrangian density is [10]
\begin{eqnarray}
{\cal L}_{B} &=& - \frac {1}{4} F^{\mu\nu}\cdot F_{\mu\nu}
+ \frac {1}{12}H^{\mu\nu\eta}\cdot H_{\mu\nu\eta}+\frac{m}{4}\varepsilon_{\mu\nu\eta\kappa}
B^{\mu\nu}\cdot F^{\eta\kappa} + B^{\mu} \cdot B_\mu  \nonumber\\
&-& \frac{i}{2}\; B^{\mu\nu} \cdot (B_1 \times F_{\mu\nu}) - (D_\mu B^{\mu\nu} - D^{\nu}\phi) \cdot B_\nu 
+ D_\mu \bar \beta \cdot D^{\mu} \beta \nonumber\\ 
&+& \frac{1}{2} \;\Bigl [(D_\mu \bar C_\nu - D_\nu \bar C_\mu ) 
- \bar C_1 \times F_{\mu\nu} \Bigr ] \cdot \Bigl [(D^\mu C^\nu - D^\nu C^\mu ) - C_1 \times F^{\mu\nu} \Bigr ]\nonumber\\
&+& \rho \cdot (D_\mu C^{\mu} - \lambda ) +  (D_\mu {\bar C}^{\mu} - \rho )\cdot \lambda, 
\end{eqnarray}
where $B_\mu = B_\mu \cdot T$ and $B_1 = B_1 \cdot T$ are the Nakanishi-Lautrup  type 
bosonic auxiliary fields and $\rho = \rho \cdot T$ and $\lambda = \lambda \cdot T$ are 
fermionic auxiliary fields. The Lorentz vector fermionic (anti-)ghost fields $(\bar C_\mu)C_\mu$ 
(with $C^2_\mu = \bar C^2_\mu= 0, C_\mu C_\nu + C_\nu C_\mu = 0, C_\mu \bar C_\nu + \bar C_\nu C_\mu = 0,$ etc.) and
bosonic (anti-)ghost fields $(\bar \beta)\beta$ are required for the unitarity in the theory and they 
carry the ghost numbers $(\mp 1)$ and $(\mp 2)$, respectively. The bosonic scalar field $\phi$ and the 
Lorentz scalar (anti-)ghost auxiliary fields $(\bar C_1)C_1$ are also 
required for the BRST invariance in the theory. It can be explicitly checked that  
\begin{eqnarray}
s_{b} {\cal L}_ {B} &=& - \partial_\mu \Bigl [\frac {m}{2}\;\varepsilon^{\mu\nu\eta\kappa} F_{\nu\eta} \cdot C_\kappa 
- \lambda \cdot B^\mu + (C_1 \times F^{\mu\nu} ) \cdot  B_\nu \nonumber\\
&-&  \rho\cdot D^\mu \beta - (D^\mu C^\nu - D^\nu C^\mu ) \cdot B_\nu  \Bigr ],
\end{eqnarray}
which shows that the action $S = \int d^4 x \;{\cal L}_B $ remains invariant under the following 
BRST symmetry transformations ($s_b$):
\begin{eqnarray}
&&s_b B_{\mu\nu} = - \; (D_{\mu}C_{\nu} -D_{\nu}C_{\mu} ) + \; C_{1} \times F_{\mu\nu},\quad
s_b C_{\mu} =- D_{\mu}\beta, \nonumber\\
&& s_b \bar C_{\mu} = B_{\mu},\quad s_b {\bar B_1} = i\; \lambda,  \quad s_b \bar C_1 = i\; B_{1}, \quad 
s_b \bar B_{\mu} = - D_{\mu} \lambda, \nonumber\\ 
&& s_b K_{\mu} =\; D_{\mu}C_{1} - C_{\mu},\quad s_b \phi = \lambda, \quad s_b C_1 = -\beta, 
\quad s_b \bar \beta = \rho, \nonumber\\ 
&& s_b [ A_{\mu}, \; F_{\mu\nu}, \; H_{\mu\nu\eta},\; \beta, \; B_1, \; \rho, \; \lambda, \; B_{\mu} ] = 0.
\end{eqnarray}
It is pertinent to point out that the above BRST transformations have been obtained from the
superfield approach to BRST formalism [10] which always produces the off-shell nilpotent $(s_b^2 = 0)$ 
BRST symmetry transformations for a given $p$-form $(p = 1, 2, 3, ...)$ gauge theory in any arbitrary
dimension.

Exploiting the basics of the Noether theorem, it turns out that the exact expression for Noether current is   
\begin{eqnarray}
J^\mu_{(b)} &=& \frac {m}{2} \; \varepsilon^{\mu\nu\eta\kappa} \; F_{\nu\eta} \cdot C_\kappa 
- \; \frac {1}{2} \; H^{\mu\nu\eta} \cdot [(D_\nu C_\eta - D_\eta C_\nu ) - C_1 \times F_{\nu\eta}]\nonumber\\
&+& [(D^\mu \bar C^\nu - D^\nu \bar C^\mu ) - \bar C_1 \times F^{\mu\nu}] \cdot (D_\nu \beta) 
+ (D^\mu \beta) \cdot \rho \nonumber\\
&+& [(D^\mu C^\nu - D^\nu C^\mu ) - C_1 \times F^{\mu\nu}] \cdot B_\nu + B^\mu \cdot \lambda.
\end{eqnarray}
The conservation of this current ($i.e.$ $\partial_\mu J^\mu_{(b)} = 0$) can be proven by exploiting 
the following set of E-L equations of motion\footnote{The present theory is highly constrained
because we have the conditions: $F_{\mu\nu} \times B^{\mu\nu} = 0, F_{\mu\nu} \times H^{\mu\nu\eta} = 0$. However,
the 2-form $F^{(2)} = d A^{(1)} + i A^{(1)} \wedge A^{(1)}$ does respect zero curvature condition
($F_{\mu\nu} = 0 $) for the choice $A^{(1)} = - i \;U\; d\; U^{-1}$ where $U \in SU(N)$.}
\begin{eqnarray}
&& D_\mu F^{\mu\nu} - \; \frac {m}{2} \; \varepsilon^{\mu\nu\eta\sigma}\; D_\mu B_{\eta\sigma} 
+ D_\mu (H^{\mu\nu\eta} \times K_\eta) 
+ \;i \; D_\mu ( B^{\mu\nu} \times B_1) \nonumber\\
&& - \;D_\mu [(D^\mu C^\nu - D^\nu C^\mu ) \times \bar C_1 ] + \;D_\mu [ (D^\mu \bar C^\nu 
- D^\nu \bar C^\mu)\times C_1]\nonumber\\
&& + \; \frac {1}{2} \; ( H^{\nu\eta\sigma} \times B_{\eta\sigma})
- ( B^{\mu\nu} \times B_\mu ) + (B^\nu \times \phi ) + (D^\nu \bar \beta \times \beta) \nonumber\\
&& + \; \bar C_\mu \times [(D^\mu C^\nu - D^\nu C^\mu ) - C_1 \times F^{\mu\nu}]  
- (\bar C^\nu \times \lambda) + \;(C^\nu \times \rho)\nonumber\\
&& + \;(D^\nu \beta \times \bar \beta ) - C_\mu \times [(D^\mu \bar C^\nu - D^\nu \bar C^\mu )
- \bar C_1 \times F^{\mu\nu} ]= 0, \nonumber\\
&& D_\mu H^{\mu\nu\eta} - \; \frac {m}{2} \; \varepsilon^{\nu\eta\kappa\sigma}\; F_{\kappa\sigma} 
- (D^\nu B^\eta - D^\eta B^\nu) 
- \; i\; (F^{\nu\eta} \times B_1) = 0,\nonumber\\
&& B_\mu = - (1/2)\; (D_\mu \phi - D^\nu B_{\nu\mu}), \;\; (B_{\mu\nu} \times F^{\mu\nu}) = 0, 
\;\; (H^{\mu\nu\eta} \times F_{\nu\eta}) = 0,\nonumber\\
&&  D_\mu (D^\mu \bar \beta ) = 0, \quad \rho = \frac {1}{2} \; (D_\mu \bar C^\mu ), 
\quad \lambda = \frac {1}{2} \; (D_\mu C^\mu ), \quad D_\mu B^\mu = 0,\nonumber\\
&& D_\mu [(D^\mu \bar C^\nu - D^\nu \bar C^\mu ) - \bar C_1 \times F^{\mu\nu}] = 
- \; D^\nu \rho, \quad D_\mu (D^\mu \rho) = 0,\nonumber\\
&& D_\mu [(D^\mu C^\nu - D^\nu C^\mu ) - C_1 \times F^{\mu\nu}] = 
- \; D^\nu \lambda, \quad D_\mu (D^\mu \lambda) = 0,\nonumber\\
&& [(D_\mu \bar C_{\nu} - D_\nu \bar C_\mu) - \bar C_1\times F_{\mu\nu}] \times F^{\mu\nu} = 0, 
\;\quad D_\mu (D^\mu \beta ) = 0,\nonumber\\
&& [(D_\mu C_\nu - D_\nu C_\mu) - C_1\times F_{\mu\nu}] \times F^{\mu\nu} = 0.
\end{eqnarray}
The above equations emerge from the Lagrangian density ${\cal L}_B.$

The conserved current (13) leads to the derivation of the conserved $(\dot Q_b = 0)$ 
and nilpotent $(Q^2_b = 0)$ BRST charge $Q_b = \int d^3x \;J^0_{(b)}$ as  
\begin{eqnarray}
Q_{b} &=& \int d^3x \Big[\frac {m}{2} \; \varepsilon^{0ijk} F_{ij} \cdot C_k 
- \frac {1}{2} \;H^{0ij} \cdot \Big(D_i C_j 
- D_j C_i - C_1 \times F_{ij}\Big)\nonumber\\
&+& \Big(D_0 \bar C_i - D_i \bar C_0 - \bar C_1 \times F_{0i}\Big) \cdot D^i \beta 
+  B^0 \cdot \lambda + (D^0 \beta) \cdot \rho \nonumber\\ 
&+& \Big(D_0 C_i - D_i C_0 - C_1 \times F_{0i}\Big) \cdot B^i\Big].
\end{eqnarray}
This charge is the generator of the BRST symmetry transformations (11).

We wrap up this section with the remarks that
(i) one can derive the BRST transformation $s_b B_{0i} = - i [B_{0i}, \; Q_b] = 
- (D_0 C_i - D_i C_0) + C_1 \times F_{0i}$ from the BRST charge (the analogue of which, we  
were unable to derive from the gauge symmetry generator $Q_v$) (cf. section 2),
(ii) one can derive {\it all} the BRST symmetry transformations for {\it all} the fields by various requirements 
(cf. section 4 below), and (iii) one is {\it not} able to derive, however, the BRST transformation 
$s_b K_\mu = D_\mu C_1 - C_\mu$ from the conserved charge $Q_b$. Thus, we conclude that, 
except for the auxiliary field $K_\mu$, all the other fields have the usual 
off-shell nilpotent ``vector'' BRST symmetry transformations.\\

\noindent
{\bf \large 4. Anti-BRST symmetry transformations and 

anti-BRST charge}\\

\noindent
In addition to ${\cal L}_B$, there is yet another generalization of ${\cal L}_{(0)}$ that includes the 
gauge-fixing and Faddeev-Popov ghost terms. Such an appropriate
(anti-BRST invariant)  Lagrangian density ${\cal L}_{\bar B}$, in its full 
blaze of glory, is  [10]
\begin{eqnarray}
{\cal L}_{\bar B} &=& - \frac {1}{4} F^{\mu\nu}\cdot F_{\mu\nu}
+ \frac {1}{12}H^{\mu\nu\eta}\cdot H_{\mu\nu\eta}+\frac{m}{4}\varepsilon_{\mu\nu\eta\kappa}
B^{\mu\nu}\cdot F^{\eta\kappa} + {\bar B}^{\mu} \cdot {\bar B}_\mu  \nonumber\\
&+& \frac{i}{2}\; B^{\mu\nu} \cdot ({\bar B}_1 \times F_{\mu\nu}) + (D_\mu B^{\mu\nu} + D^{\nu}\phi) \cdot {\bar B}_\nu 
+ D_\mu \bar \beta \cdot D^{\mu} \beta \nonumber\\ 
&+& \frac{1}{2} \;\Bigl [(D_\mu \bar C_\nu - D_\nu \bar C_\mu ) 
- \bar C_1 \times F_{\mu\nu} \Bigr ] \cdot \Bigl [(D^\mu C^\nu - D^\nu  C^\mu ) - C_1 \times F^{\mu\nu} \Bigr ]\nonumber\\
&+& \rho \cdot (D_\mu C^\mu - \lambda ) +  (D_\mu {\bar C}^\mu - \rho )\cdot \lambda.
\end{eqnarray}
The above Lagrangian density respects the off-shell nilpotent $(s_{ab}^2 = 0)$ anti-BRST symmetry transformations 
$s_{ab}$ as listed below 
\begin{eqnarray}
&&s_{ab} B_{\mu\nu} = - \; (D_{\mu}\bar C_{\nu} -D_{\nu} \bar C_{\mu} ) + \; \bar C_{1} \times F_{\mu\nu},
\quad s_{ab} \bar C_{\mu} = - D_{\mu} \bar \beta, \nonumber\\
&& s_{ab} C_{\mu} = \bar B_{\mu}, \quad s_{ab} B_{\mu} = D_{\mu} {\rho},
\quad s_{ab} C_1 = i\; \bar B_{1},\quad s_{ab} \phi = -\rho, \nonumber\\
&& s_{ab} \bar C_1 = - \bar \beta, \quad s_{ab} B_{1} = - i \; \rho,
\quad s_{ab} K_{\mu} =\; D_{\mu} \bar C_{1} - \bar C_{\mu}, \nonumber\\ 
&& s_{ab} \beta = - \lambda, 
\quad s_{ab} [A_{\mu}, \;F_{\mu\nu}, \;H_{\mu\nu\eta}, \;\bar \beta, \;  \bar B_1, \; 
\rho, \; \lambda, \; \bar B_{\mu} ] = 0,
\end{eqnarray}
because ${\cal L}_{\bar B}$ transforms to a total spacetime derivative as 
\begin{eqnarray}
s_{ab} {\cal L}_{\bar B} &=& - \partial_\mu \;\Bigl [ \frac {m}{2}\;\varepsilon^{\mu\nu\eta\kappa} 
F_{\nu\eta} \cdot \bar C_\kappa + \rho \cdot {\bar B}^\mu  
- ({\bar C}_1 \times F_{\mu\nu} ) \cdot {\bar B}_\nu \nonumber\\
&+& \lambda \cdot D^\mu {\bar \beta} + (D^\mu {\bar C}^\nu - D^\nu {\bar C}^\mu ) \cdot {\bar B}_\nu  \Bigr ].
\end{eqnarray}
As a consequence, the action $S = \int d^4x \;{\cal L}_{\bar B} $ remains invariant under $s_{ab}$.

A few noteworthy points are in order. First, under the (anti-)BRST transformations, the 
kinetic terms $(- \frac {1}{4}\; F^{\mu\nu} \cdot F_{\mu\nu} $
and $\frac {1}{12}\; H^{\mu\nu\eta} \cdot H_{\mu\nu\eta} )$, owing their origin to the exerior derivative 
$d = dx^\mu \partial_\mu$ (with $d^2 = 0$), remain invariant. Second, the Nakanishi-Lauturp type auxiliary 
fields $\bar B_\mu$ and $\bar B_1$, introduced in (15), are constrained to satisfy the Curci-Ferrari 
(CF) type restrictions as given below
\begin{eqnarray}
B_\mu + \bar B_\mu = - D_\mu \phi, \qquad B_1 + \bar B_1 = i \phi.
\end{eqnarray}
It should be recalled that, the fields $B_\mu$ and $B_1$, were introduced in the definition of the 
BRST invariant Lagrangian density ${\cal L}_B$. Third, the above CF-type of restrictions have been derived 
from the superfield approach to BRST formalism for the dynamical non-Abelian 
2-form gauge theory [10]. These are (anti-)BRST invariant as it can be checked that 
$s_{(a)b} [B_\mu + \bar B_\mu + D_\mu \phi] = 0 , \;s_{(a)b} [B_1 + \bar B_1 - i \phi] = 0$
where $s_b B_1 = 0,   s_b \bar B_1 = i \lambda,  s_b B_\mu = 0,  s_b \bar B_\mu = - D_\mu \lambda,  
s_{ab} \bar B_1 = 0, s_{ab} B_1 = - i \rho, s_{ab} \bar B_\mu = 0, s_{ab} B_\mu = D_\mu \rho$. 
Fourth, it can be checked that $s_{(a)b}$ obey off-shell nilpotency  $(s^2_{(a)b} = 0)$ and absolute 
anticommutativity ($i.e.$ $s_s s_{ab} + s_{ab}s_b = 0$)  if we exploit appropriately the CF-type 
conditions (18). Fifth, both the Lagrangian density ${\cal L}_B$  and ${\cal L}_{\bar B}$ are coupled 
and equivalent (due to CF-type conditions (18)) as it can be checked that {\it both} of them respect 
the (anti-)BRST symmetry transformations. This statement can be corroborated by the following observations,
namely;
\begin{eqnarray}
s_{ab} {\cal L}_{B} &=& - \partial_\mu \Big[\frac {m}{2} \; \varepsilon^{\mu\nu\eta\kappa} 
F_{\nu\eta} \cdot \bar C _\kappa + (D^\mu \bar \beta) \cdot \lambda + \bar B^\mu \cdot \rho 
+ B^{\mu\nu} \cdot (D_\nu \rho)\nonumber\\
&+&\Big \{(D^\mu \bar C^\nu - D^\nu \bar C^\mu ) - \bar C_1 \times F^{\mu\nu} \Big\} \cdot (D_\nu \phi 
+ \bar B_\nu) \Big] \nonumber\\
&+& \frac{i}{2} \;\Big[ (D^\mu \bar C^\nu - D^\nu \bar C^\mu ) 
- \bar C_1 \times F^{\mu\nu}\Big ] \cdot \Big( B_1 
+ \bar B_1 - i \phi\Big)\times F_{\mu\nu}\nonumber\\
&+& D_\mu \Big[ (D^\mu \bar C^\nu - D^\nu \bar C^\mu ) - \bar C_1 \times F^{\mu\nu}\Big] 
\cdot \Big(B_\nu + \bar B_\nu + D_\nu \phi \Big)\nonumber\\
&+& (D_\mu \rho) \cdot \Big(B^\mu + \bar B^\mu + D^\mu \phi \Big),
\end{eqnarray}
\begin{eqnarray}
s_{b} {\cal L}_{\bar B} &=& - \partial_\mu \Big[\frac {m}{2} \; \varepsilon^{\mu\nu\eta\kappa} 
F_{\nu\eta} \cdot C _\kappa - (D^\mu \beta) \cdot \rho - B^\mu \cdot \lambda 
+ B^{\mu\nu} \cdot (D_\nu \lambda)\nonumber\\
&-&\Big \{(D^\mu C^\nu - D^\nu C^\mu ) - C_1 \times F^{\mu\nu} \Big\} \cdot (D_\nu \phi 
+ B_\nu) \Big] \nonumber\\
&-& \frac{i}{2} \;\Big[ (D^\mu C^\nu - D^\nu C^\mu ) - C_1 \times F^{\mu\nu}\Big ] \cdot \Big( B_1 
+ \bar B_1 - i \phi\Big)\times F_{\mu\nu}\nonumber\\
&-& D_\mu \Big[ (D^\mu C^\nu - D^\nu C^\mu ) - C_1 \times F^{\mu\nu}\Big] \cdot \Big(B_\nu 
+ \bar B_\nu + D_\nu \phi \Big)\nonumber\\
&-& (D_\mu \lambda) \cdot \Big(B^\mu + \bar B^\mu + D^\mu \phi \Big).
\end{eqnarray}
The above equations, in addition to (10) and (17), establish the equivalence of ${\cal L}_B$ 
and ${\cal L}_{\bar B}$ as far as the validity of CF-type restrictions and the existence of the nilpotent
(anti-)BRST symmetries are concerned.

The infinitesimal continuous anti-BRST symmetry transformations (16) lead to the derivation 
of Noether current $J^\mu_{(ab)}$ as  
\begin{eqnarray}
J^\mu_{(ab)} &=& \frac {m}{2} \; \varepsilon^{\mu\nu\eta\kappa} \; F_{\nu\eta} \cdot \bar C_\kappa 
- \; \frac {1}{2} \; H^{\mu\nu\eta} \cdot [(D_\nu \bar C_\eta - D_\eta \bar C_\nu ) 
- \bar C_1 \times F_{\nu\eta}]\nonumber\\
&-& [(D^\mu \bar C^\nu - D^\nu \bar C^\mu ) - \bar C_1 \times F^{\mu\nu}] \cdot \bar B_\nu 
- (D^\mu \bar \beta) \cdot \lambda \nonumber\\
&-& [(D^\mu C^\nu - D^\nu C^\mu ) - C_1 \times F^{\mu\nu}] \cdot (D_\nu \bar \beta) - \bar B^\mu \cdot \rho.
\end{eqnarray}
The conservation law $\partial_\mu J^\mu_{(ab)} = 0$ can be proven by exploiting the E-L 
equations of motion derived from the Lagrangian density ${\cal L}_{\bar B}$. In fact, many equations of motion are 
common for the Lagrangian density ${\cal L}_B$ and ${\cal L}_{\bar B}$. The ones that are different 
from (13) and derived from ${\cal L}_{\bar B}$ are
\begin{eqnarray}
&& D_\mu F^{\mu\nu} - \; \frac {m}{2} \; \varepsilon^{\mu\nu\eta\sigma}\; D_\mu B_{\eta\sigma} 
+ D_\mu (H^{\mu\nu\eta} \times K_\eta) - \;i \; D_\mu (B^{\mu\nu} \times \bar B_1 ) \nonumber\\
&& - \;D_\mu [(D^\mu C^\nu - D^\nu C^\mu )\times \bar C_1 ] + \;D_\mu [(D^\mu \bar C^\nu 
- D^\nu \bar C^\mu )\times C_1]\nonumber\\
&& + \; \frac {1}{2} \; (H^{\nu\eta\sigma}\times B_{\eta\sigma})
+ (B^{\mu\nu}\times \bar B_\mu) + (\bar B^\nu \times \phi ) + (D^\nu \bar \beta \times \beta ) \nonumber\\
&& + \; \bar C_\mu \times [(D^\mu C^\nu - D^\nu C^\mu ) - C_1 \times F^{\mu\nu}] 
- (\bar C^\nu \times \lambda) + \;(C^\nu \times \rho)\nonumber\\
&& + \;(D^\nu \beta \times \bar \beta ) - C_\mu \times [(D^\mu \bar C^\nu - D^\nu \bar C^\mu )
- \bar C_1 \times F^{\mu\nu} ]= 0, \nonumber\\
&& D_\mu H^{\mu\nu\eta} - \; \frac {m}{2} \; \varepsilon^{\nu\eta\kappa\sigma}\; F_{\kappa\sigma} 
+ (D^\nu \bar B^\eta - D^\eta \bar B^\nu) 
+ \; i\; (F^{\nu\eta} \times \bar B_1) = 0,\nonumber\\
&& \bar B_\mu = - \;(1/2)\; (D_\mu \phi + D^\nu B_{\nu\mu}), \qquad D_\mu \bar B^\mu = 0.
\end{eqnarray}
The conserved current $J^\mu_{(ab)}$ leads to the derivation of the generator $Q_{ab} = \int d^3x \;J^0_{(ab)}$ 
of the anti-BRST symmetry transformations (16) as 
\begin{eqnarray}
Q_{ab} &=& \int d^3x \Big[\frac {m}{2} \; \varepsilon^{0ijk} F_{ij} \cdot \bar C_k 
- \frac {1}{2} \;H^{0ij} \cdot \Big(D_i \bar C_j 
- D_j \bar C_i - \bar C_1 \times F_{ij}\Big)\nonumber\\
&-& \Big(D_0 C_i - D_i C_0 - C_1 \times F_{0i}\Big) \cdot (D^i \bar \beta) -  \bar B^0 \cdot \rho 
- (D^0 \bar \beta) \cdot \lambda \nonumber\\ 
&-& \Big(D_0 \bar C_i - D_i \bar C_0 - \bar C_1 \times F_{0i}\Big) \cdot \bar B^i\Big].
\end{eqnarray}
The above charge is conserved $(\dot Q_{ab} = 0)$ and off-shell nilpotent of order two ($i.e.$ $Q^2_{ab} = 0$).
The latter property establishes the fermionic nature of $Q_{ab}$.

A few comments are in order. It can be readily seen that $Q_{ab}$ generates the anti-BRST symmetry 
transformations for the field $B_{0i}$ as $s_{ab} B_{0i} = - i [B_{0i}, \; Q_{ab}] = 
- (D_0 \bar C_i - D_i \bar C_0) + \bar C_1 \times F_{0i}.$ Furthermore, it is interesting to point out that 
$Q_b Q_{ab} + Q_{ab}Q_b = 0$ if and only if the CF-type restrictions (18) are 
exploited for its proof. Finally, the conserved and nilpotent charge $Q_{ab}$ is unable to generate the 
anti-BRST symmetry transformation 
for the {\it special} auxiliary field $K_\mu$
($i.e.$ $s_{ab} K_\mu = D_\mu \bar C_1 - \bar C_\mu$). This is a novel observation.

The most surprising thing is 
that the above {\it specific} transformation {\it can not} be derived even from the requirements of (i) the
(anti-)BRST invariance of CF-type restrictions (18), (ii) the nilpotency property, and 
(iii) the anticommutativity property of $s_{(a)b}$ 
($i.e.$ $s_b s_{ab} + s_{ab}s_b = 0$). This is {\it a new} observation in the
context of the application of BRST approach to topologically  massive 4D non-Abelian theory 
(which is drastically different from the application of the same approach to its Abelian counterpart (see, 
$e.g.,$ [6,7])). \\

\noindent
{\bf \large 5. Ghost symmetry, ghost charge and BRST algebra}\\

\noindent
The ghost part of the Lagrangian density of the theory
\begin{eqnarray}
{\cal L}_g &=& \frac{1}{2} \;\Bigl [(D_\mu \bar C_\nu - D_\nu \bar C_\mu ) 
- \bar C_1 \times F_{\mu\nu} \Bigr ] \cdot \Bigl [(D^\mu C^\nu - D^\nu C^\mu ) 
- C_1 \times F^{\mu\nu} \Bigr ]\nonumber\\
&+& D_\mu \bar \beta \cdot D^\mu \beta + \rho \cdot (D_\mu C^{\mu} - \lambda ) 
+  (D_\mu {\bar C}^{\mu} - \rho )\cdot \lambda,
\end{eqnarray}
remains invariant under the following scale transformations
\begin{eqnarray}
&& C_1 \to e^{+\Sigma} \;C_1, \quad \bar C_1 \to e^{- \Sigma}\; \bar C_1, 
\quad C_\mu \to e^{+ \Sigma} \; C_\mu, \quad \bar C_\mu \to e^{- \Sigma} \;\bar C_\mu,\nonumber\\
&& \beta \to e^{+2 \;\Sigma} \;\beta, \quad  \bar \beta \to e^{- 2 \;\Sigma}\; \bar \beta, 
\quad \rho \to e^{- \Sigma}\; \rho, \quad s_g \;\lambda = e^{+\Sigma}\; \lambda, \nonumber\\
&& (A_\mu,  B_{\mu\nu},  B_\mu,  \bar B_\mu,  B_1,  \bar B_1,  \phi) \to (A_\mu,  B_{\mu\nu},  
B_\mu,  \bar B_\mu,  B_1,  \bar B_1,  \phi),
\end{eqnarray}
where $\Sigma$ is a global scale parameter and numbers in the exponential denote the corresponding 
ghost number of the fields ($e.g.$ the ghost field  $\beta$ has the ghost number equal to +2). It is elementary 
to check that the following infinitesimal version of the above scale symmetry transformations 
\begin{eqnarray}
&& s_g C_1 = \Sigma \;C_1, \quad s_g \bar C_1 = - \Sigma\; \bar C_1, \quad s_g C_\mu = \Sigma \; C_\mu, 
\quad s_g  \bar C_\mu = - \Sigma \;\bar C_\mu,\nonumber\\
&& s_g \beta = 2 \;\Sigma \;\beta, \quad s_g \bar \beta = - 2 \;\Sigma\; \bar \beta, 
\quad s_g \rho = - \Sigma\; \rho, \quad s_g \lambda = \Sigma\; \lambda, \nonumber\\
&& s_g [A_\mu, \; B_{\mu\nu}, \; B_\mu, \; \bar B_\mu, \; B_1, \; \bar B_1, \; \phi] = 0,
\end{eqnarray}
leads to the derivation of the conserved Noether current
\begin{eqnarray}
J^\mu_{(g)} &=& 2 \beta \cdot D^\mu \bar \beta - 2 \bar \beta \cdot D^\mu \beta 
- C_\nu \cdot[(D^\mu \bar C^\nu - D^\nu \bar C^\mu ) - \bar C_1 \times F^{\mu\nu}] \nonumber\\
&-& \bar C_\nu \cdot [(D^\mu C^\nu - D^\nu C^\mu) - C_1 \times F^{\mu\nu}] - C^\mu \cdot \rho 
- \bar C^\mu \cdot \lambda.
\end{eqnarray}
The conservation law $\partial_\mu J^\mu_g = 0$ can be proven by exploiting the E-L equation of 
motion derived from the ${\cal L}_B$ and ${\cal L}_{\bar B}$ (cf. (13), (22)) for the ghost,
anti-ghost and other appropriate fields of the theory.

The conserved ghost 
charge $Q_g = \int d^3x \;J^0_{(g)}$, derived from the Noether conserved current $J^\mu_{(g)}$, is as follows 
\begin{eqnarray}
Q_g &=& \int d^3x \Big[2 \beta \cdot D^0 \bar \beta - 2 \bar \beta \cdot D^0 \beta 
- C_i \cdot[(D^0 \bar C^i - D^i \bar C^0 ) - \bar C_1 \times F^{0i}] \nonumber\\
&-& \bar C_i \cdot [(D^0 C^i - D^i C^0) - C_1 \times F^{0i}] - C^0 \cdot \rho - \bar C^0 \cdot \lambda \Big].
\end{eqnarray}
The above charge generates\footnote{It is obvious that $Q_g$ does not generate the 
ghost transformation for $\rho, \lambda, C_1$ and $\bar C_1$ which are auxiliary fields.
These transformations are derived from other considerations.} the infinitesimal version of (26). 
Using the definition of a generator 
($e.g.$ $s_b Q_b = - i \{ Q_b, \;Q_b\},  s_b Q_g = - i [ Q_g, \; Q_b] $, $etc.$), it can be seen that the 
following standard BRST algebra emerges, namely; 
\begin{eqnarray}
&Q^2_b = 0, \quad Q^2_{ab} = 0, \quad \{Q_b, \; Q_{ab}\} = Q_b \;Q_{ab} + Q_{ab} \;Q_b = 0,&\nonumber\\
&i \;[Q_g,\; Q_b] = + \; Q_b, \qquad i\; [Q_g,\; Q_{ab}] = - \;Q_{ab}.& 
\end{eqnarray}
It should be pointed out that, for the proof of the anticommutativity of the $Q_b$ and $Q_{ab}$ 
($i.e.$ $\{Q_b, \; Q_{ab}\} = 0$), we have to exploit the beauty and strength of the CF-type restrictions 
in (18). It is trivial to note, in passing, that the ghost
number of $Q_{(a)b}$ is $(\mp 1)$. As a consequence, the transformations, generated by $Q_{(a)b}$,
decrease/increase the ghost number of the fields by one.\\

\noindent
{\bf \large 6. Conclusions}\\

\noindent
In our present investigation, we have concentrated on the ``vector" gauge symmetry transformations of the 4D 
topologically massive non-Abelian gauge theory and exploited it in the context of BRST analysis. We have shown that 
{\it all} 
the fields of the present theory have proper ($i.e.$ off-shell nilpotent and absolutely anticommuting\footnote{To 
prove the absolute anticommutativity $(s_b s_{ab} + s_{ab} s_b = 0)$, one has to invoke the CF-type
restrictions (18) that emerge in the superfield approach to BRST formalism [10].}) (anti-)BRST transformations 
(cf. (11), (16)). All these transformations have been tapped in the derivation of the conserved, nilpotent and 
anticommuting (anti-)BRST charges $Q_{(a)b}$ (cf. (14), (23)).

As it turns out, the generators $Q_{(a)b}$ are able to produce all the 
nilpotent (anti-)BRST symmetry transformations for the 
basic fields of the theory. Such transformations for the auxiliary fields are, as usual, derived from 
the requirements of the nilpotency and anticommutativity of the (anti-)BRST symmetry transformations. The 
(anti-)BRST invariance of the CF-type restrictions also plays a key role in such an endeavor. In our present
work there exists a very {\it special} auxiliary field, however. It transpires that 
the generators $Q_{(a)b}$ and nilpotency as well as 
anticommutativity  requirements are unable to produce the nilpotent (anti-)BRST transformations of 
$K_\mu$ field. This is a new observation in the application of BRST formalism to our present theory.

It may be mentioned, at this stage, that we have exploited the ``scalar" gauge symmetry transformation for BRST 
analysis in our earlier work [11] where we have found that the (anti-)BRST charges are {\it not} able
to generate the (anti-)BRST symmetry transformation for $B_{0i}$ and $K_\mu$ fields. This should be contrasted with our
present investigation where $Q_{(a)b}$ are unable to produce the (anti-)BRST symmetry transformations for 
{\it only} the 
auxiliary field $K_\mu$. Thus, there is a key difference between the novel features that emerge from the BRST
analysis of the ``scalar" and ``vector'' gauge symmetries.

It is worthwhile to point out that the construction of a renormalizable, unitary and consistent 4D non-Abelian 2-form
gauge theory is an outstanding problem which is yet to be resolved completely. The Freedmen-Townsend (FT) model [2] and 
Lahiri model [3,4] are two of the quite well-known models which have their own virtues and vices. In a recent 
paper [15], one of us has shown the novel features of the FT model within the framework of BRST formalism. The central 
goal of our studies [10,11,15], including the present one, is to understand these models [2-4] clearly and, if possible, propose a model which is free of 
all the drawbacks of the above models. \\

\noindent
{\bf Acknowledgements}\\

\noindent
One of us (RK) is grateful to UGC, Govt. of India, for financial support. It is a great pleasure
for both of us to acknowledge very useful communications with A. Lahiri on certain aspects of
our present investigation. We are also thankful to our esteemed referee for some clarifying 
and enlightening remarks.\\

\begin{center}
{\large \bf Appendix}
\end{center}
Here we provide a brief synopsis of the similarities between Lahiri model [3,4] and the 
celebrated Proca model within the framework of St{\"u}ckelberg formalism (see, e.g. [16] for a review). 
It can be noticed that if we express the Lagrangian density (1) 
in terms of the redefined field $\tilde B_{\mu\nu}$, viz.;
\begin{eqnarray}
\tilde B_{\mu\nu} = B_{\mu\nu} + (D_\mu K_\nu - D_\nu K_\mu),
\end{eqnarray}
the auxiliary field $(K_\mu)$ disappears from the curvature tensor $H_{\mu\nu\eta}.$ Furthermore, 
the redefined topological mass term of the Lagrangian density (1):
\begin{eqnarray}
\frac{m}{4}\; \varepsilon^{\mu\nu\eta\kappa}\; \tilde B_{\mu\nu} \cdot F_{\eta\kappa} 
&=& \frac{m}{4}\; \varepsilon^{\mu\nu\eta\kappa}\; 
B_{\mu\nu} \cdot F_{\eta\kappa} + \partial_\mu \Big[\frac{m}{2}\; 
\varepsilon^{\mu\nu\eta\kappa}\; K_\nu \cdot F_{\eta\kappa}\Big]\nonumber\\
&-& \frac{m}{2}\varepsilon^{\mu\nu\eta\kappa}\; K_\nu \cdot \Big(D_\mu F_{\eta\kappa}\Big),
\end{eqnarray}
remains invariant
(modulo a total spacetime derivative term) because of the validity of the Bianchi identity 
($D_\mu F_{\nu\eta} + D_\nu F_{\eta\mu} + D_\eta F_{\mu\nu} = 0$). Thus, the auxiliary field $K_\mu$ 
is completely eliminated from the whole theory. As a consequence, even though the theory respects the 
``scalar" (i.e. Yang-Mills) gauge symmetry transformations, it fails to respect the ``vector" (i.e. tensor) gauge
symmetry transformations. This observation is, however, true for all the theories 
where the St{\"u}ckelberg trick is applied.

The above key observation should be contrasted with the Proca model. The Lagrangian density 
of the latter is as follows
\begin{eqnarray}
{\cal L}_P = - \frac{1}{4} \;F_{\mu\nu}F^{\mu\nu} + \frac{m^2}{2} \; A_\mu A^{\mu},
\end{eqnarray}
where $F_{\mu\nu} = \partial_\mu A_\nu - \partial_\nu A_\mu$ and $m$ is the mass of the 
Abelian 1-form gauge field $A_\mu$.
With the inclusion of the St{\"u}ckelberg field $\phi$ (by the application of the
standard technique), the above Lagrangian
density becomes [16]
\begin{eqnarray}
{\cal L}_S = - \frac{1}{4} \;F_{\mu\nu}F^{\mu\nu} + \frac{m^2}{2} \; A_\mu A^{\mu} 
+ \frac{1}{2}\; \partial_\mu \phi\; \partial^\mu \phi + m A_\mu \partial^\mu \phi.
\end{eqnarray}
It is well-known fact that the above Lagrangian density 
(33) remains invariant under the following gauge transformations ($\delta_g$)
\begin{eqnarray}
\delta_g A_\mu = \partial_\mu \Lambda (x), \;\;\qquad \;\;\;\delta_g \phi = - \;m \;\Lambda (x),
\end{eqnarray} 
where $\Lambda (x)$ is a local infinitesimal transformation parameter. It can be checked 
that, if we incorporate the following redefinition:
\begin{eqnarray}
\tilde A_\mu  = A_{\mu} - \frac{1}{m}\;\partial_\mu \phi, 
\end{eqnarray}
in the above Lagrangian density,
the St{\"u}ckelberg field $\phi$ is totally eliminated from ${\cal L}_S$. 
The ensuing theory, thus, does not respect the gauge transformation $\delta_g A_\mu = \partial_\mu \Lambda$. 
We conclude that the Lahiri model of 
the 4D non-Abelian 2-form theory (with the auxiliary field $K_\mu$) is 
exactly same, in structure, as the Proca theory with the St{\"u}ckelberg field $\phi$.
Thus, the auxiliary field $K_\mu$ of the Lahiri model is the  St{\"u}ckelberg field in the 
true sense of the word.

In contrast to the above St{\"u}ckelberg's tricks [16], the spontaneous symmetry breaking, though the
Higgs mechanism, is yet another technique to generate a mass for the gauge field. For instance, 
one knows that in the polar decomposition of the scalar field, the Goldstone mode is eliminated 
due to the $U(1)$ gauge transformation and the photon field acquires a mass in the 
{\it Abelian} Higgs model 
(see, e.g. [17] for details). The final theory, with the mass term for the photon, does not respect, however, the 
local $U(1)$ gauge symmetry transformations for obvious reasons. We lay stress on the fact that
the St{\"u}ckelberg trick and the SSB of the gauge symmetries, through the Higgs mechanism, are entirely 
different techniques to generate the mass.

\end{document}